\documentclass{PoS}
\usepackage{mcite}

\title{Exploring the chiral regime of $N_f=2$ QCD with mixed actions}

\ShortTitle{Mixed action computations on fine dynamical lattices}

\author{\speaker{F. Bernardoni}\footnote{After July 1st 2010: DESY, Platanenallee 6, 15738 Zeuthen}, Pilar Hern\'andez\\%
        IFIC - Edificio Institutos de Investigaci\'on\\
			Apartado de Correos 22085\\
			E-46071 Valencia - Espa\~na \\
        E-mail: \email{Fabio.Bernardoni@desy.de, pilar.hernandez@ific.uv.es}}

\author{Nicolas Garron\\    
	University of Edinburgh, School of Physics and Astronomy\\
	James Clerk Maxwell Building\\
	Mayfield Road\\
	Edinburgh
   E-mail: \email{ngarron@staffmail.ed.ac.uk}}

\author{Silvia Necco\\
        CERN\\
        CERN CH-1211\\
			Gen\`eve 23\\
			Switzerland
        E-mail: \email{necco@mail.cern.ch}}

\author{Carlos Pena\\
        Dpto. de F\'isica Te\'orica and Instituto de F\'isica Te\'orica UAM/CSIC\\ Universidad Aut\'onoma de Madrid\\
         Cantoblanco E-28049 Madrid\\
          Spain 
        E-mail: \email{carlos.pena@uam.es}}

\abstract{We report on our simulations with Neuberger valence fermions on CLS $N_f=2$ configurations with non-perturbatively $O(a)$-improved Wilson sea quarks. We consider the matching of QCD to ChPT in the so called mixed-regime in which the sea quarks are in the $p$-regime while the valence quarks are in the $\epsilon$-regime. From this matching, we can get information on $\Sigma$, $L_6$ and the combination $L_8+2L_6+2L_7$.
\vspace{2cm}
\begin{flushright}
IFT-UAM/CSIC-10-76, FTUAM-10-30 \\
IFIC/10-46\\
CERN-PH-TH/2010-264 
\end{flushright}
}

\FullConference{The XXVIII International Symposium on Lattice Filed Theory\\
		 June 14-19,2010\\
		 Villasimius, Sardinia Italy}

\begin{document}

\section{Introduction}

In this work we want to test the matching between Lattice QCD and ChPT in a special finite volume regime in which the valence quarks have masses $m_{\rm v}$ in the $\epsilon$-regime (that is $m_{\rm v}\Sigma V \lesssim 1$, but $L \Lambda_\chi \gg1$ where $\Lambda_\chi$ is the cutoff of ChPT, $\Sigma$ is the quark condensate, $V=L^3\times T$ is the volume). As discussed in~\cite{Gasser:1987ah}, under these conditions slow pion modes dominate the path integral in the effective theory, so the expansion in powers of $M_\pi^2/\Lambda_\chi^2 \sim p^2/\Lambda_\chi^2$ breaks down and a new power-counting has to be used. In particular the pion zero modes have to be treated non perturbatively. Nevertheless, the low energy constants (LECs) that appear in the ChPT Lagrangian are unaffected by finite size effects so that this regime provides a framework in which the LECs can be determined with different systematics with respect to usual ``infinite volume'' studies. 

Our aim is to use a mixed action approach, in which chiral symmetry is exactly preserved at the level of valence quarks only. We consider overlap valence quarks on top of $N_f=2$ CLS ensembles,\footnote{https://twiki.cern.ch/twiki/bin/view/CLS/WebHome} obtained from simulations with non-perturbatively $O(a)$-improved Wilson sea quarks. The use of cheaper Wilson quarks allows us to consider rather small lattice spacings ($a \sim 0.08 fm$) and sufficiently large physical volumes ($L \sim 2 fm$). We also take a Partially Quenched approximation in which the sea quark masses $m_s$ stay in the $p$-regime, that is they satisfy $m_s\Sigma V \gg1$. A consistent power counting for this regime was introduced in~\cite{Bernardoni:2007hi,*Bernardoni:2008ei}.

The use of overlap in the valence sector is clearly suitable to study low energy QCD phenomenology, where the effects of spontaneous breaking of chiral symmetry are more visible. Moreover it allows us to define unambiguosly the topological charge $\nu$ through the index theorem. As a first application we have matched the dependence of $\langle \nu^2 \rangle$ on the sea quark mass to the prediction of NLO ChPT in the mixed regime, that depends on the LECS: $\Sigma$ and a combination of $L_6$, $L_7$ and $L_8$. 

In the $\epsilon$-regime at a fixed $\nu$, the LO partition function of ChPT has been shown to be equivalent to a random matrix theory (RMT) \cite{Shuryak:1992pi}, where many analytical predictions can be obtained for spectral quantities, such as the spectral density or the distribution of individual eigenvalues of the Dirac operator. Here we show that ChPT in the mixed regime, up to NLO, can also be matched to a RMT. The dependence on the sea quark mass of the parameter of the RMT can be extracted from this matching and can be shown to depend on $\Sigma$ and $L_6$.

A detailed paper has been published recently \cite{Bernardoni:2010nf}, to which we refer for  a more extensive discussion.

\section{Matching QCD in the mixed-regime with Random Matrix Theory}
\label{sec:chpt}

As argued in \cite{Bernardoni:2007hi}
, the heavier $p$-regime sea quarks behave as decoupling particles that can be integrated out. The dependence on $m_s$ can then be reabsorbed in the LECs. The matching to a RMT requires the further integration out of the non-zero momentum modes in ChPT.

For a theory with $N_f$ quarks in the $\epsilon$-regime the partition function of ChPT at fixed $\nu$, ${\cal Z}_\nu^{ChPT}({\cal M})$ can be written at LO as:
\begin{equation}
{\cal Z}_\nu^{ChPT}({\cal M})=\int \,d\xi  \int_{U(N_f)} \left[d U_0\right]   \det (U_0)^{\nu}  \exp\left( \frac{\Sigma V}{2}{\rm Tr} \left[ \mathcal{M} (U_0+U_0^{\dagger})\right] \right) e^{-\int d^4x\,\mbox{Tr} [\partial_\mu \xi \partial_\mu \xi]}
\end{equation}
where ${\mathcal M} = \delta_{ij} m_i$ is the $N_f \times N_f$ mass matrix, $U_0$ and $\xi$ parametrize respectively the non perturbative zero modes and the perturbative non zero modes of the pions. In finite volume the momentum is quantized 
and for sufficiently low energies the non zero modes can be decoupled, order by order in the $\epsilon$-expansion if the Pseudo-Goldstone masses $M_{ab}^2 \equiv \frac{\Sigma (m_a+m_b)}{F^2}$ satisfy $M_{ab} \ll L^{-1}$. This leads to the definition of a Zero Modes Chiral Theory (ZMChT) with partition functional 
\begin{equation}
{\cal Z}_{\nu,\,\epsilon}^{ZMChT}({\cal M}) \propto \int_{U(N_f)} \left[d U_0\right]   \det (U_0)^{\nu}  \exp\left( \frac{\Sigma V}{2}{\rm Tr} \left[ \mathcal{M} (U_0+U_0^{\dagger})\right] \right)\,.
\end{equation}
In \cite{Shuryak:1992pi}, it was shown that ${\cal Z}_\nu^{ZMChT}({\cal M})$ is equivalent to the partition functional of a Random Matrix Theory (RMT) of matrices $W$ of size $N$, that depends only on the number of flavours, $N_f$, and the corresponding mass parameters $\hat{m}$, viz. 
\begin{equation}
{\cal Z}_{\nu,\,\epsilon}^{ZMChT}({\cal M})\simeq {\rm RMT}_N\{N_f,\hat{m}\}\,,
\label{eq:shuryak}
\end{equation} 
and there must be an identification $N \hat{m}_i = m_i \Sigma V \equiv \mu_i$.
From this relation spectral correlation functions of the Dirac operator can be related to the corresponding quantities computed  in RMT. In particular, if $x$  is the $k$-th smallest eigenvalue of the matrix $\sqrt{W^\dagger W}$, the probability distribution associated to the microscopic eigenvalue $\zeta=Nx$, $p_k^\nu(\zeta;\{\mu\})$ can be calculated in RMT. Then also the distribution of individual low-lying eigenvalues of the massless Dirac operator can be predicted \cite{Nishigaki:1998is} from the equivalence Eq.~(\ref{eq:shuryak}) in terms of just one parameter, the chiral condensate $\Sigma$.

This relation has been tested in the quenched approximation 
\cite{Edwards:1999ra,*Bietenholz:2003mi,*Giusti:2003gf}, respectively in $N_f=2$ and $N_f=2+1$ dynamical simulations in \cite{DeGrand:2006nv,*Lang:2006ab,*Fukaya:2007yv} and \cite{Hasenfratz:2007yj,*Fukaya:2009fh}. 

In the mixed regime with $N_v$ $\epsilon$-regime quarks ($M_{vv} \ll L^{-1}$) and $N_s$ $p$-regime quarks ($L^{-1} \sim M_{ss}$), at low energies all non-zero modes and also the zero modes of the heavier pions must be integrated out. A ZMChT is obtained
corresponding to $N_f= N_v$ flavours as opposed to the total $N_v+N_s$ that one would obtain if all the quarks were in the $\epsilon$-regime. 

At LO the ZMChT in the mixed-regime is simply ChPT without the heavy modes (the integration over them gives an irrelevant normalization factor):
\begin{eqnarray}
\left. Z_{\nu,\,m}^{ZMChT} \right|_{LO} \propto \int_{U(N_{\rm v})} [d U_0] (\det U_0 )^\nu \exp\left({\Sigma V\over 2} {\rm Tr}\left[  {\mathcal M}_{\rm v} \left(U_0 + U^\dagger_0\right)\right]\right).
\label{eq:zmchtlo}
\end{eqnarray} 
where $ {\mathcal M}_{\rm v} $ is the mass matrix $ {\mathcal M}$ projected to the valence sector. According to the eq.~(\ref{eq:shuryak}), this partition function is then equivalent to an $N_{\rm v}$  RMT. In particular, it is important to stress that the ZMChT has $N_{\rm v}$ flavours, while the full ChPT from which it is derived corresponds to $N_f=N_{\rm v}+N_s$ flavours.  In particular, for $N_{\rm v} \rightarrow 0$,  the ZMChT or RMT we expect to find is the {\it quenched} one, while the couplings should be those of an $N_f=N_s$ theory. \\
The matching at NLO still does not modify the structure of the ZMChT theory. The dynamics of heavy modes can be absorbed in the couplings appearing at LO in eq.~(\ref{eq:zmchtlo}), that is $\Sigma$. The integrations over the perturbative modes result in a change  $\Sigma \rightarrow \Sigma_{\rm eff}$, which for degenerate quarks and in the limit $N_{\rm v} \to 0$, reads:
\begin{eqnarray}
\Sigma_{\rm eff}(m_s) &=& \Sigma\, \Bigg\{ 1+ { 2 m_s \Sigma  \over F^4} \Bigg[ {\beta_2 \over 2} + {1 \over 16\pi^2} \log(M_\rho V^{1/4}) + 32 L^r_6(M_\rho)
-{1 \over 16 \pi^2} \log\left({m_s \Sigma \over F^2 M_\rho^2}\right)\Bigg]
\nonumber\\
&&~~~~~~
- \frac{\beta_1}{2F^2V^{1/2}} -{2 \over F^2} g_1\left(\sqrt{{\Sigma m_s\over F^2}}, L, T\right)\Bigg\}\,,
\label{eq:sigmaeff}
\end{eqnarray}
where $M_\rho = 770$ MeV is the $\rho$ meson mass, while the function $g_1(M, L, T)$ expressing the volume effects on the propagator and the $\beta_i$ are defined in \cite{Hasenfratz:1989pk}. In our case $T/L=2$, $\beta_1=0.08360$ and $\beta_2=-0.01295$.
The quenched limit of the zero-mode integral over $U(N_{\rm v})$ is calculated in~\cite{Splittorff:2002eb} and matches quenched RMT (qRMT). \\
Comparing the eigenvalues computed numerically in this Partially Quenched setup with the predictions of qRMT, we can extract  $\Sigma_{\rm eff}$ of eq.~(\ref{eq:sigmaeff}):  
\begin{equation}\label{rmt_matching}
\langle \zeta_k \rangle^\nu_{\rm qRMT}   =\left. \Sigma_{\rm eff}(M_{ss})\right|_{N_{\rm v}=0} V \langle\lambda_k\rangle ^\nu_{\rm QCD}(M_{ss})     ,
\qquad
\langle \zeta_k \rangle^\nu_{\rm qRMT} = \int \zeta\, p_k^\nu(\zeta;0) \zeta.
\end{equation}
On the other hand, the prediction for the ratios $\langle \zeta_k \rangle^\nu_{\rm qRMT}/\langle \zeta_l \rangle^\nu_{\rm qRMT}$ is parameter-free and can be compared directly with $\langle\lambda_k\rangle ^\nu_{\rm QCD}/\langle\lambda_l\rangle ^\nu_{\rm QCD}$ at any fixed $M_{ss}$.

The distribution of topological charge  is, on average, controlled by the sea quarks only. In the case when $N_{\rm v} \rightarrow 0$ and the $N_s$ sea quarks are degenerate, we found:
\begin{eqnarray}
\langle \nu^2\rangle = {m_s \Sigma V\over N_s} & &\left[ 1-  {N_s^2 -1 \over N_s} \left({\Sigma m_s\over 8 \pi^2 F^4} \log\left( {2 m_s\Sigma \over F^2M_\rho^2}\right) + g_1(\sqrt{{2 m_s\Sigma \over F^2}},L,T)\right) \right.\nonumber\\
&+& \left. {32 \Sigma m_s \over F^4} \left( L^r_8(M_\rho)+ N_s L^r_6(M_\rho) + N_s L^r_7(M_\rho)  \right) \right]\,.
\label{eq:nu2nlo}
\end{eqnarray}
This result agrees with \cite{Mao:2009sy} in infinite volume and with \cite{Aoki:2009mx} in the $p$-regime. Note the appearance of  $L_7^r(M_\rho)$,  for which no prediction has yet been obtained on the lattice. \\

\section{Results on Dirac spectral observables}
\label{sec:num}

We have carried out our computations on the CLS lattices of size $48\times 24^3$ named D$_4$, D$_5$ and D$_6$\cite{DelDebbio:2006cn}.
The configurations have been generated with non-perturbatively $O(a)$-improved Wilson fermions at $\beta=5.3$ using the DD-HMC algorithm and saved configurations are separated by 30 HMC trajectories of length $\tau=0.5$. The lattice spacing has been determined to be $a=0.0784(10)~$fm in \cite{DelDebbio:2006cn}, which implies that our lattices have physical size $L\simeq 1.88$ fm and sea pion masses of 426, 377 and 297 MeV, respectively. However, preliminary results from other determinations (eg using the mass of the $\omega$  baryon)  yield $a\simeq 0.070~Ê$fm \cite{small:a}. We will consider both values in our analysis.

On these configurations we have built the massless Neuberger-Dirac operator $D_{\rm N}$ \cite{Neuberger:1997fp} with the parameter $s$ that governs the locality fixed to $s=0.4$. We have computed $\nu$ using the index theorem $\nu=n_+ -n_-$ where $n_+$ ($n_-$) is the number of zero modes of $D_{\rm N}$ with positive (negative) chirality, and the 10 lowest eigenvalues $\gamma$ of $D_{\rm N}$. Given that the $\gamma$ appear in general in complex conjugated pairs and lie on a circle in the complex plane $\gamma=\frac{1}{\overline{a}}(1-e^{i\phi})$, we have used the projection $\lambda=\sqrt{\gamma\gamma^*}=\frac{1}{\overline{a}}\sqrt{2(1-\cos\phi)}$ to compare with RMT. Expectation values were computed at fixed absolute value of the topological charge $|\nu|$.

\begin{figure}
\begin{center}
\includegraphics[width=7cm]{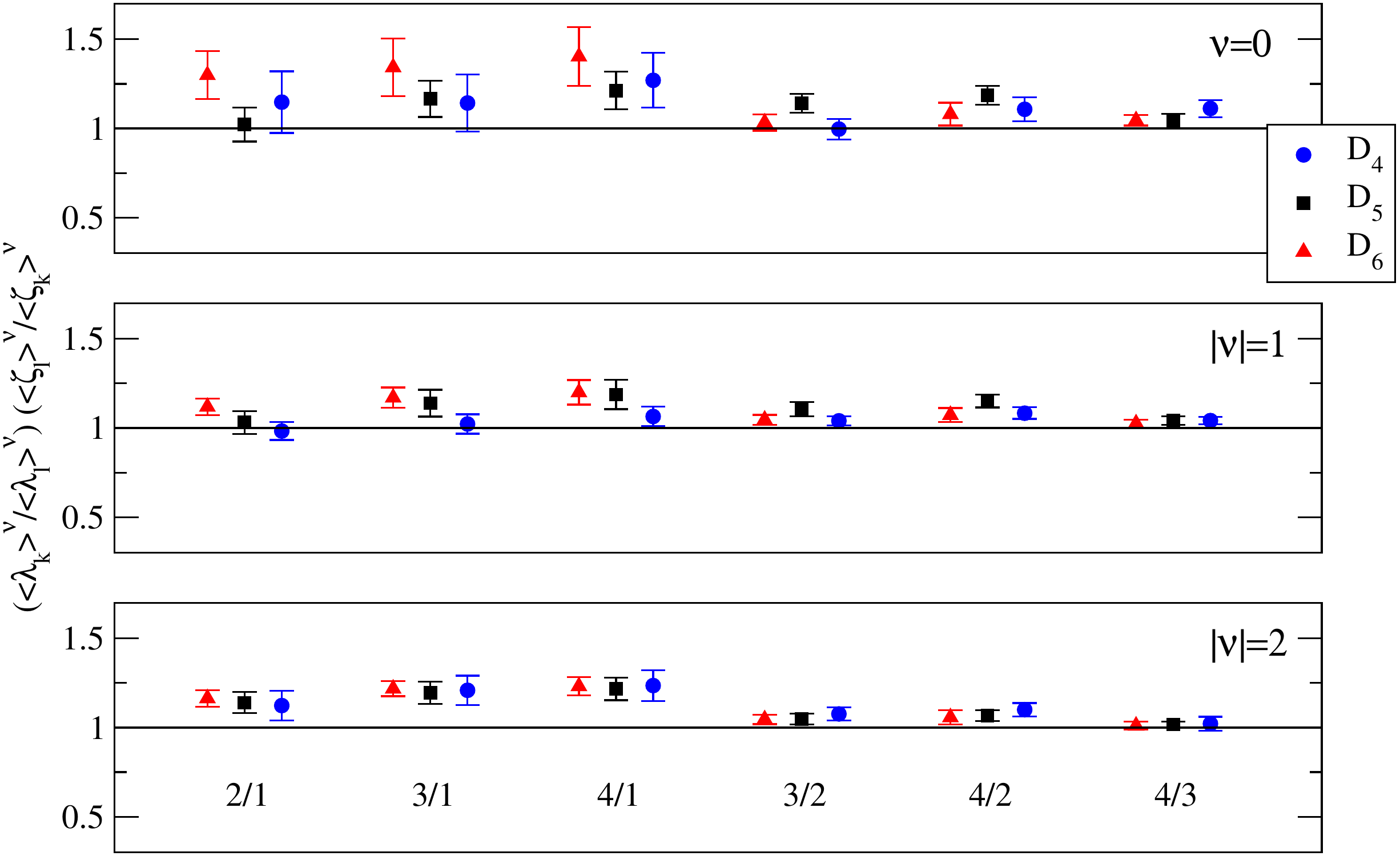}\includegraphics[width=7cm]{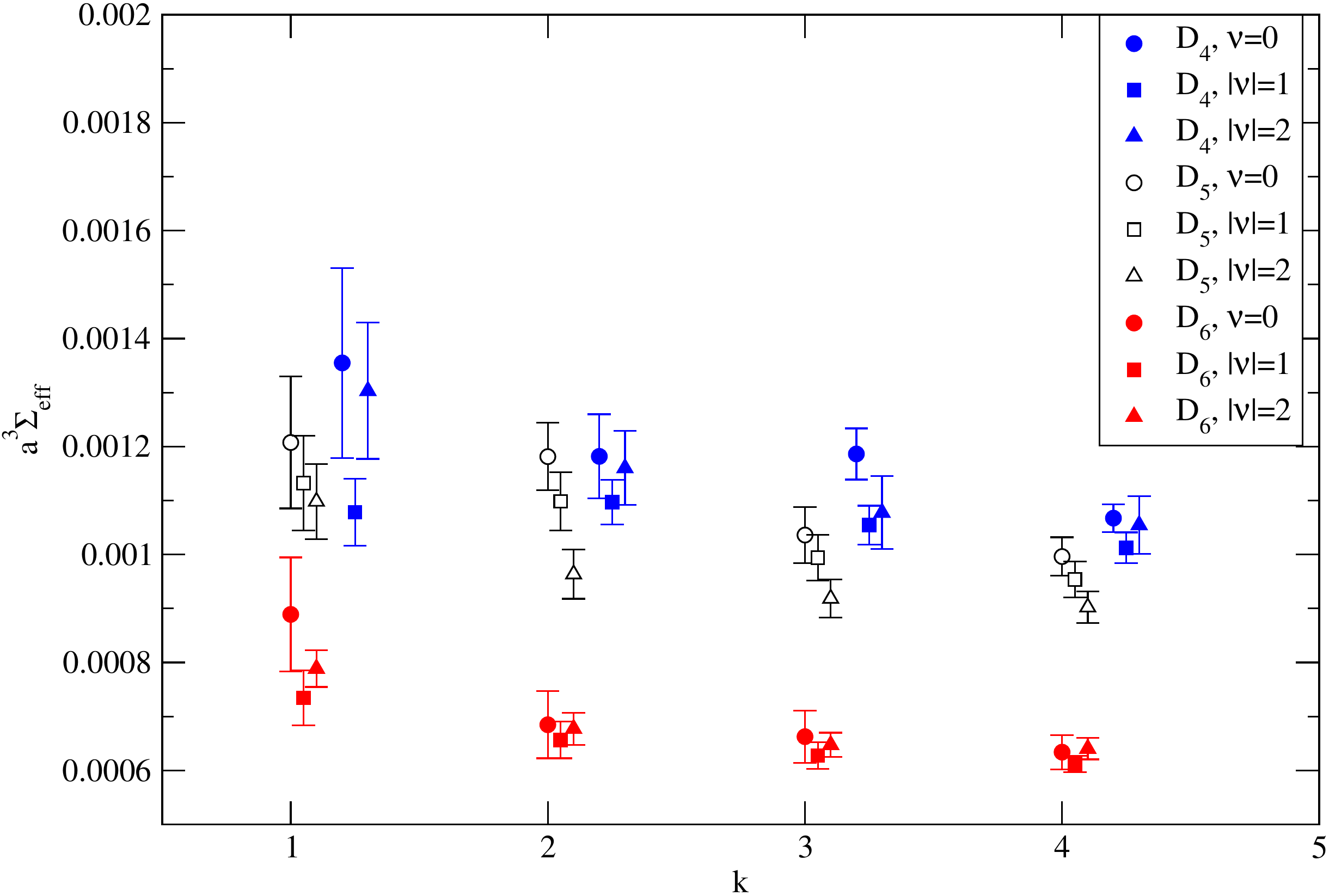}
\caption{Left: eigenvalue ratios at fixed $|\nu|$, normalized to qRMT predictions, for the indices $(k,l)=(2,1), (3,1), (4,1), (3,2), (4,2), (4,3)$. Right: $\Sigma_{\rm eff}$ bare extracted from the matching in eq.~(\protect\ref{sigmaeff_mat}), for $k=1,2,3,4$ and $|\nu|=0,1,2$. The data for D$_4$ have been shifted in the horizontal axis for better clarity.}\label{fig:ratios}
\end{center}
\end{figure}

In Fig. \ref{fig:ratios} (left) we report the ratios $\langle\lambda_k\rangle^{\nu}/\langle\lambda_l\rangle^{\nu}$ normalized to the corresponding qRMT predictions, for $|\nu|=0,1,2$ and for several combinations $k,l$ given at the bottom of the plot. This allows to appreciate the precision and level of agreement with qRMT of each specific case.
While the RMT prediction seems to work well for ratios not involving $\lambda_1$, the ratios $\langle\lambda_k\rangle/\langle\lambda_1\rangle$ exhibit somewhat more significant deviations.  

In the spirit of the mixed-regime ChPT analysis, our data also allow to study
the mass dependence of the effective condensate, cf. eq.~(\ref{eq:sigmaeff}).
The values of the bare effective condensate extracted from the matching
\begin{equation}\label{sigmaeff_mat}
\Sigma_{\rm eff}(M_{ss})=\frac{\langle \zeta_k \rangle^\nu_{\rm qRMT}}{V \langle\lambda_k\rangle ^\nu_{\rm QCD}}
\end{equation}
for $k=1,2,3,4$ and $|\nu|=0,1,2$ are shown in Fig. \ref{fig:ratios} (right), where one can 
observe that, at fixed value of the sea quark mass, $\Sigma_{\rm eff}$ does not depend on $k$ and $\nu$ within the statistical precision (with larger errors for $k=1$). We average over  $k=2,3,4$ and $|\nu|=0,1,2$ and estimate systematic effects by adding the values for $k=1$ in the average.

\section{Fits to NLO Chiral Perturbation Theory}

On the basis of the evidence presented in the previous section, now we assume that the matching to ChPT in the mixed-regime works in the range of masses and volumes we used, and  try to extract the low-energy couplings from the sea-quark mass dependence of the two quantities $\Sigma_{\rm eff}$ and $\langle\nu^2\rangle$.

\begin{figure}
\begin{center}
\includegraphics[width=7cm]{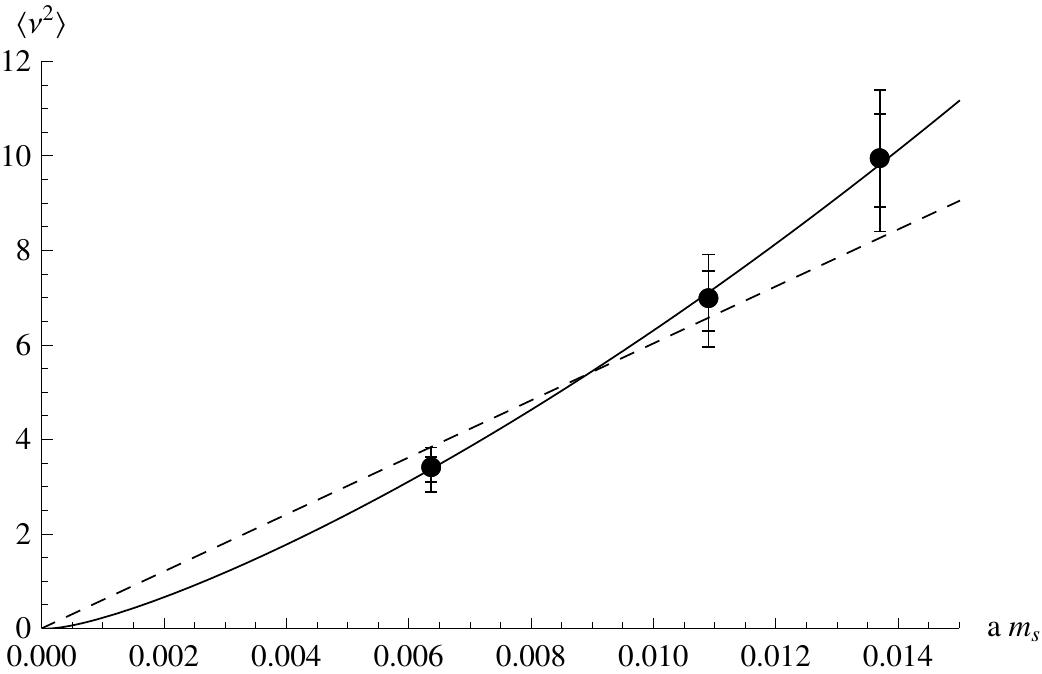}\includegraphics[width=7cm]{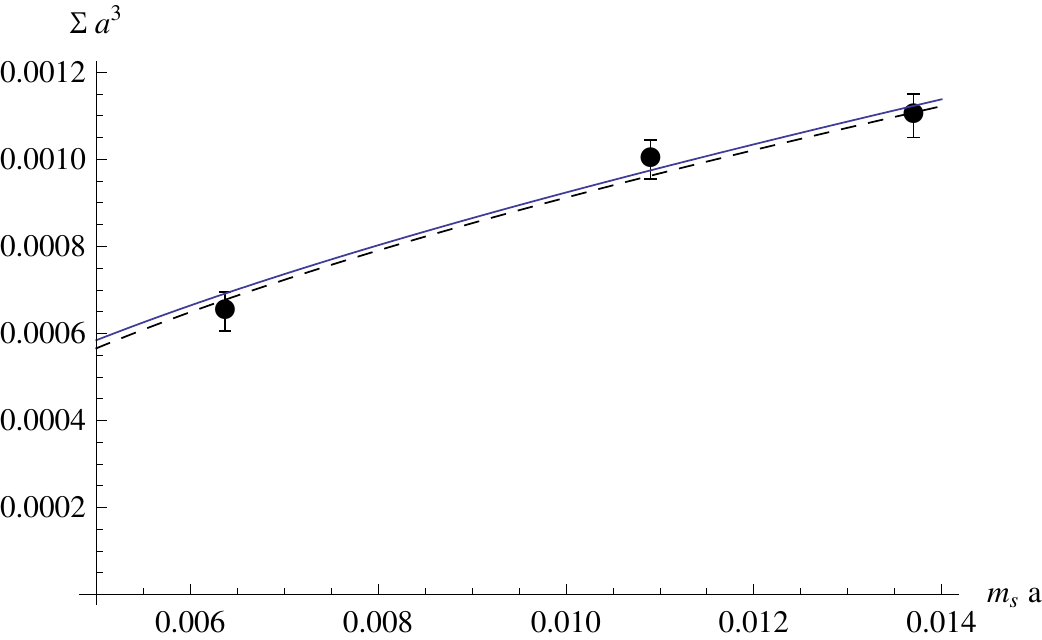}
\caption{Left: $\langle \nu^2\rangle$ versus the sea quark mass $a m_s^{\overline{\rm MS}}(2{\rm GeV})$. The smaller errors are the statistical ones and the largest include our estimate of autocorrelations. The dashed and solid lines correspond to the best LO and NLO ChPT fits respectively (for $a=0.078$~fm). Right: $\Sigma_{\rm eff} a^3$. The dashed and solid lines correspond to the best fit for $\Sigma$ and $L_6^r(M_\rho)$ at $F=90~$MeV,  taking the scale to be $0.070~$fm and $0.078~$fm respectively.}
\label{fig:nu2}
\end{center}
\end{figure}

We first consider the topological charge distribution. The statistical error in this quantity is fairly large, but it is encouraging to see that there is a very clear dependence on the sea quark mass (see Fig.~\ref{fig:nu2} left). We have fitted both to the full NLO formula in eq.~(\ref{eq:nu2nlo}), and to the linear LO behaviour. In either case $m_s$ is taken to be the PCAC Wilson mass renormalized in the $\overline{\rm MS}$ scheme at 2~GeV. The results for D$_4$ and D$_5$ are taken from \cite{DelDebbio:2007pz}, while we have computed that of D$_6$.

At NLO we fit for $\Sigma$ and the combination $\left[L^r_8 + 2 (L^r_6+L^r_7)\right](M_\rho)$. The value of $F$ is taken to be 90~MeV; the systematic uncertainty related to this choice is estimated by varying $F$ by $\pm 10$~MeV.
In physical units we get for $a=0.078-0.070~$fm:
\begin{eqnarray}
\Sigma^{\overline{\rm MS}}(2\; {\rm GeV}) & = &\left[262^{(33)(4)}_{(34)(5)} ~Ê{\rm MeV}\right]^3 \;- \; \left[287^{(35)(5)}_{(36)(7)} ~Ê{\rm MeV}\right]^3\\
\left[L^r_8 + 2 (L^r_6+L^r_7)\right](M_\rho) &= & 0.0018(30) \;- \; 0.0023(43)
\end{eqnarray}
where the first error is coming out of the fit and the second is the effect of changing $F$. Since our ensembles are not large enough to allow a precise estimate of the effect of autocorrelations, we quote errors computed with a conservative approach \cite{Bernardoni:2010nf}.

Let us now turn to $\Sigma_{\rm eff}$. In this case, the dependence on $m_s$ is expected starting at NLO in ChPT. We perform a two-parameter NLO fit, where we fix $F$ and fit for $\Sigma$ and $L^r_6(M_\rho)$. \\
In order to obtain the scalar density renormalization factor in the $\overline{\rm MS}$ scheme for the valence overlap fermions we apply the matching condition
\begin{eqnarray}
\left. \left(Z^{\overline{\rm MS}}_{\rm S}\right)^{-1}  m \right|^{\rm overlap}_{M^{\rm ref}_\pi} =   \left. m^{\overline{\rm MS}}(2~\mbox{GeV})\right|^{\rm Wilson}_{ M^{\rm ref}_\pi}
\end{eqnarray}
at $aM_{\rm ref}=aM_{ss}$. We obtain $Z^{\overline{\rm MS}}_{\rm S}(2~\mbox{GeV}) = 1.84(10),
$ where the error is dominated by the one in $aM_{vv}$, i.e. in the determination of
the unitary point.
Obviously, several checks need to be done to ensure that this result is robust, such as checking the dependence on the reference pion mass, as well as on the sea quark mass.

With this estimate for the renormalisation factor and fixing $F=90$ MeV, the result we obtain from the fit is, for $a=0.078-0.070~$fm 
\begin{eqnarray}
\Sigma^{\overline{\rm MS}}(2\; {\rm GeV})  &=& \left[255^{(12)(1)}_{(13)(4)} ~Ê{\rm MeV}\right]^3 \;- \; \left[280^{(13)(4)}_{(14)(5)} ~Ê{\rm MeV}\right]^3\\  
L_6^r(M_\rho) &=& 0.0010(6)\;-\;0.0015(10) \,, 
\end{eqnarray}
where the only systematic error that has been estimated is that associated to the change of $F$ by $\pm 10$~MeV. \\
Although the fit is good (see Fig.~\ref{fig:nu2} right), it would be desirable to have more sea quark masses and smaller statistical errors to assess the systematics. Particularly useful would be to test the finite-size scaling. 

This value of $\Sigma$  is consistent with the one obtained from the topological susceptibility above, and  both are in nice agreement with  the alternative determination of \cite{Giusti:2008vb}, that extracted the condensate from the spectrum of the Wilson-Dirac operator on  $N_f=2$ CLS configurations at the same lattice spacing and sea quark masses, but in a larger physical volume. \\

\section{Conclusions}

Using a mixed action approach, together with being in the mixed regime and in the PQ approximation has allowed us to match the Dirac low spectrum with the ChPT predictions with significantly finer lattices than in previous studies. Our determination of $\Sigma$ lies in the same ballpark as many of these previous determinations. More work is however required to quantify the systematic uncertainties involved in this method: autocorrelations, systematics in the chiral fit and finite $a$. \\


\providecommand{\href}[2]{#2}\begingroup\raggedright\endgroup


\end{document}